\documentclass[prl,twocolumn,showpacs,amsmath,amssymb]{revtex4}

\usepackage{graphicx}% Include figure files
\usepackage{dcolumn}% Align table columns on decimal point
\usepackage{bm}% bold math
\usepackage{ifthen}

\begin{document}

\title{Metal-nonmetal transition and excitonic ground state in InAs/InSb
quantum dots}

\author{Lixin He}
%\affiliation{National Renewable Energy Laboratory, Golden, Colorado 80401}
\author{Gabriel Bester}
%\affiliation{National Renewable Energy Laboratory, Golden, Colorado 80401}
\author{Alex Zunger}
\affiliation{National Renewable Energy Laboratory, Golden, Colorado 80401}
%\email{lhe@nrel.gov}
%\homepage{sst.nrel.gov}
\date{\today{}}% It is always \today, today,

\begin{abstract}
Using atomistic pseudopotential and
configuration-interaction many-body calculations,
we predict a metal-nonmetal transition and an excitonic ground state in the
InAs/InSb quantum dot (QD) system.
For large dots, the conduction band minimum of the InAs dot lies below the
valence band maximum of the InSb matrix. Due to quantum confinement,
at a critical size calculated 
here for various shapes, the single-particle gap $E_g$ becomes very
small. Strong electron-hole correlation effects are induced by the spatial
proximity of the electron and hole wavefunctions, and by the lack of strong (exciton unbinding)
screening, afforded by the existence of fully discrete 0D confined energy
levels. These correlation effects overcome $E_g$, leading to 
the formation of a bi-excitonic ground state
(two electrons in InAs and two holes in InSb)
being energetically more favorable (by $\sim$ 15 meV) than
the state without excitons.  
We discuss the excitonic phase transition on QD arrays in the low dot
density limit.
  
\end{abstract}

% 71.35.Lk Collective effects 
%(Bose effects, phase space filling, and excitonic phase transitions) 
%73.21.-b Electron states and collective excitations in multilayers, 
%quantum wells, mesoscopic, and nanoscale systems  
%(for electron states in nanoscale materials, see 73.22.-f)
%
% 73.21.La Electron states and collective excitations in multilayers, quantum
%          wells, mesoscopic, and nanoscale system: Quantum dots
% 78.67.Hc Optical properties of low-dimensional structures: Quantum dots
% 73.22.-f Electronic structure of nanoscale materials: clusters,
%          nanoparticles, nanotubes, and nanocrystals
% 71.15.-m computational methodology use 71.15.-m
%nanoparticles, nanotubes, and nanocrystals

\maketitle

The formation of excitons in semiconductors and insulators 
usually requires energy,
e.g. photons, for one has to excite carriers across the single-particle
band-gap $E_g$. 
There is a special interest, however, in the possibility of forming excitons 
exothermically,
i.e. an ``excitonic ground state'' as envisioned by Mott~\cite{mott61} 
and Keldysh 
{\it et al.}~\cite{keldysh}. Indeed, the electron-hole system exhibits a 
rich range of phases
\cite{halperin68,littlewood02}
as a function of the carrier density and effective-mass ratio $m_e/m_h$,
including various excitonic insulating states such as  molecular solid, 
exciton liquid, Mott insulator, and also various metallic phases. 
The excitonic ground state 
is of fundamental interest in itself because excitons
can be a better alternative to atoms for studying Bose-Einstein 
condensation\cite{bec_book,butov02} on account of the lighter
excitonic mass, thus higher condensation temperature. 
It is natural to search for excitonic ground states in systems 
where $E_g$ is small, yet the screening is weak enough so as to 
prevent unbinding of the exciton. 
The search 
in {\it bulk solids} \cite{bucher91} has thus
focused on {\it indirect} gap semiconductors and
semimetals to reduce screening, but excitonic 
ground states have not been conclusively 
observed so far in such systems.
Ground state excitons were also searched in {\it nanostructures}, specifically
in spatially indirect quantum-wells~\cite{datta85,zhu95}, where
electrons and holes are confined in different spatial regions.
In ``type II'' systems such as GaInAs/InP or CdTe/CdSe, 
electrons are localized in the well, 
whereas holes are localized on the barrier, so screening is weak,  
but $E_g$ is finite. 
In contrast, in ``type III'' heterostructures such as~\cite{magri00} 
InAs/GaSb, the conduction band minium (CBM) 
of the InAs well is lower than the valence band maximum (VBM) of the GaSb
barrier, 
so at certain a well thickness one can have $E_g$$ \rightarrow$0~\cite{magri00}, 
as well as separation of electrons from holes.
Thus, at this thickness, one could expect an excitonic
ground state if the electron-hole correlation energy will be large enough to
stabilize the complex. 
Recent experiments~\cite{cheng95} show evidence
for existing excitonic ground state in type III InAs/Al$_{x}$Ga$_{1-x}$Sb 
quantum-well superlattices; however, the binding energy was small 
(estimated at $\sim$ 3-4 meV) and strong magnetic fields are needed to
stabilize the system.
This is because in a 2D periodic well or superlattice,
extended states exist in the in-pane direction
leading to rather weak excitonic binding. 

The advantage of a type-III system can, however, be utilized without the
disadvantage of 2D periodicity underlying a quantum-well, 
if one considers a type III {\it 0D quantum-dot} (QD).
The well-known InAs/Al$_{x}$Ga$_{1-x}$Sb epitaxial system~\cite{cheng95} is
inappropriate here, since it is lattice-matched, so dots will not form in a
strain-induced Stranski-Krastanov (SK) growth~\cite{bimberg_book}. 
We propose here a new dot
matrix system, InAs/InSb, which has a type-III band alignment\cite{magri02}, 
and a lattice mismatch, and is  
hence amenable to epitaxial SK growth. 
Using an empirical pseudopotential approach~\cite{zunger01}, we
find that as the dot size is reduced, electron levels localized on InAs move up
in energy, whereas hole states, localized at the interface, 
initially above
the electron levels, move down in energy. The system reaches a degeneracy
$E_g\sim$0 of electron and hole single-particle levels at a critical
size, predicted here for realistic dot shapes. As a result of the 0D confinement
of both electrons and holes, there is but a small number of fully discrete
bound states, so screening is limited. 
Using many-body configuration-interaction (CI),
we find that consequently at the 
critical size, the exciton is bound by as much as $\sim$15 meV
$\gg$ $E_g$, thus forming an excitonic ground state. 
Our study characterizes theoretically the properties of the excitonic ground
state in this system, offering experimentally testable predictions.

{\it Identifying the material system}:
The customary way of selecting a {\it strained} dot/matrix system
\cite{bimberg_book} is to require that the dot material has a smaller 
gap than the matrix material (e.g., InAs/GaAs or CdSe/ZnSe) so that all
carriers can be confined inside the dot.
In conventional semiconductors\cite{lb22a} the
requirement $E_g^{\rm dot} < E_g^{\rm matrix}$ implies
$a^{\rm dot} > a^{\rm matrix}$, where $a$ is the
bulk lattice constant, leading to compressive strain. 
Here, however, we seek $E_g$$\rightarrow$0 with spatially separated, yet
confined carriers,
so we might need a system where the CBM of the dot material is {\em
below} the VBM of the matrix material. This could be met when the dot material
has a {\it larger} band gap (thus, a smaller lattice constant) than the matrix
material, leading to {\it tensile} strain. SK growth under tensile strain was
demonstrated before for PbTe/PbSe\cite{pinczolits98}. 
The InAs/InSb system~\cite{magri00} has a lattice mismatch of $\sim$ 7\%, and 
the CBM of InAs is below
the VBM of InSb by 0.17 eV for the unstrained system.
This guarantees that the electrons will be confined on InAs, and holes on
InSb. We will see below that one can further chose a strained dot {\it shape} that
will bring the holes to the interface, thus providing proximity, yet spatial
separation of electrons and holes.

\begin{figure}
\includegraphics[width=2.8in,angle=0]{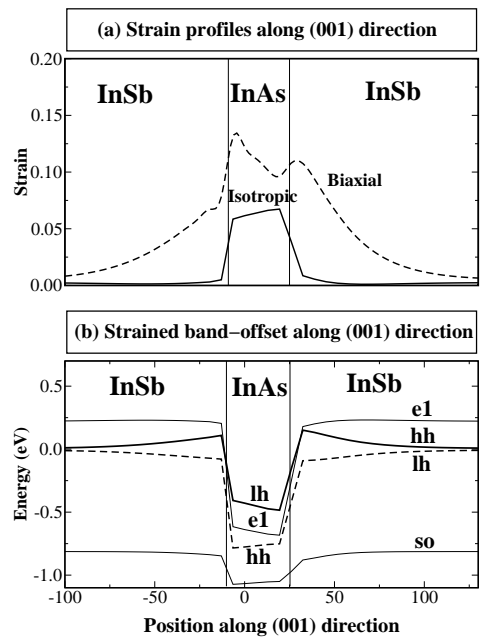}
\caption{ (a) The isotropic and biaxial strain profiles along 
the (001) direction; 
(b) Strain-modified band-offset along (001) direction calculated from
Pikus-Bir model. {\bf e1}, {\bf hh}, 
{\bf lh} and {\bf so} denote the band characters of the confining potential.
}
\label{fig:strain-offset}
\end{figure}

{\it Strain and confinement}:
We have tested various dot shapes and sizes. Here we show 
three lens-shaped InAs dots of base/height 
=104/26, 143/36 and 156/39 \AA\; 
labeled as D1, D2 and D3, respectively, as well as spherical dots, all
embedded in an InSb matrix.
The supercell that
includes the dot and the matrix contains up to about 6$\times$10$^5$ atoms. 
The strain is relaxed by minimizing the strain-energy 
with respect to the displacement of all atoms, representing the energy as
a sum of bond-bending and bond-stretching force constants (the valence force
field method (VFF)~\cite{keating66}). 
Denoting by $\epsilon_{\alpha\beta}$, the $\alpha\beta$-component
of the strain tensor,
Fig.~\ref{fig:strain-offset}(a) shows the strain profile along the (001)
direction across the dot D2,
demonstrating that the isotropic strain 
$I$=$\epsilon_{xx}$+$\epsilon_{yy}$+$\epsilon_{zz}$
is positive (tensile), almost a constant inside 
the dots and decays to zero very rapidly
in the barrier, whereas the biaxial component $B=\left[
(\epsilon_{xx}-\epsilon_{yy})^2 + (\epsilon_{yy}-\epsilon_{zz})^2
+ (\epsilon_{zz}-\epsilon_{xx})^2 \right]^{1/2}$ is nonzero and nonconstant
inside the dot, but decays slowly to zero outside.  
To observe how this strain modifies the 
electronic properties,
we show in Fig. \ref{fig:strain-offset}(b) the
strain-modified confining potentials obtained by inserting the strain of Fig. 
\ref{fig:strain-offset}(a) into the Pikus-Bir equations\cite{pikus61}. 
%We see in Fig. \ref{fig:strain-offset}(b) that
The isotropic strain lowers both the CBM and the
VBM of InAs, but the lowering of CBM is greater, leading to a large 
reduction of the gap from the unstrained bulk value of 0.42 eV to 0.01 eV.
Very importantly, because of the curved shape of the strained dot 
leading to a large biaxial strain at the interface,
the heavy-hole potential 
has a maximum at the InAs/InSb {\it interface}. 
We thus expect (and find below) that
hole states will localize at the interface, rather than be extend throughout the
barrier. This will enhance electron-hole binding.

%------------------------------------------------

{\it Method of calculation}: We relax the atomic position,
$\{{\bf R}_{n,\alpha}\}$ via the VFF method and 
construct the total pseudopotential of the system 
$V({\bf r})$ by superposing the
local, screened atomic pseudopotential $v_{\alpha}({\bf r})$
of all (dot+matrix) atoms and the nonlocal
spin-orbit potentials $V_{\rm SO}$: 
$V({\bf r}) = V_{\rm SO} + \sum_{n,\alpha} v_\alpha
({\bf r}-{\bf R}_{n,\alpha})$. 
We use the Linear Combination of Bulk Bands (LCBB) 
method~\cite{wang99b}, where the Hamiltonian 
$-1/2\nabla^2 + V({\bf r})$ is diagonalized in a basis
$\{\phi_{m,\epsilon,\lambda}({\bf k})\}$ of Bloch orbitals of 
band index $m$
and wave vector ${\bf k}$ of material $\lambda$ (= InAs, InSb, GaAs), strained
uniformly to strain $\epsilon$.
This LCBB approach~\cite{wang99b} produces accurate 
results for many nanostructures~\cite{bester03}, and greatly surpasses in
accuracy the ${\bf k} \cdot {\bf p}$  method 
which limits the basis to $m$=VBM+CBM at k=0
only. The pseudopotentials of Ref.~\onlinecite{magri02} are used for InAs/InSb
with minor modifications~\cite{insb_pp_note}.
Many-body effects are included via the configuration interaction (CI)
method\cite{franceschetti99} 
by expanding the total wavefunction
in Slater determinants for single and bi-excitons formed from all
of the confined single-particle 
electron and hole states. The
Coulomb and exchange integrals are computed numerically from the
pseudopotential single-particle states, 
using the microscopic
position-dependent dielectric constant~\cite{franceschetti99}. 
Our CI approach is very similar to the
Bethe-Salpeter equations\cite{onida02}, except that
in the latter case all the exchange integrals are unscreened,
whereas since the exchange potential has 
a long-range component~\cite{franceschetti99},
we do allow its screening.

\begin{figure}
\includegraphics[width=2.8 in,angle=0]{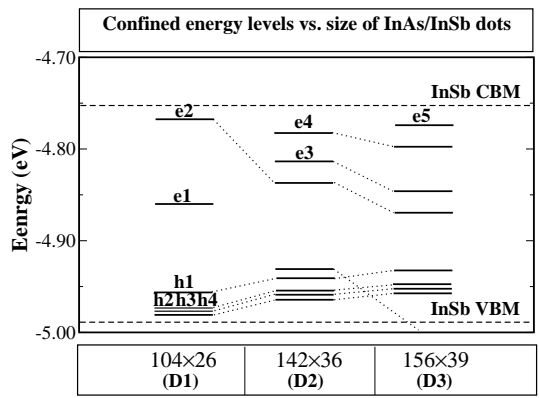}
\caption{ Single-particle spectrum of D1 (104 \AA $\times$ 26 \AA)
, D2 (142 \AA $\times$ 36 \AA) , D3 (156 \AA $\times$ 39 \AA).
e1, e2, e3, e4 are the InAs-confined 
electron states, while h1, h2, h3, h3 are
localized hole states.}
\label{fig:energy}
\end{figure}
%----------------------------------------------

\begin{figure}
\includegraphics[width=2.8in,angle=0]{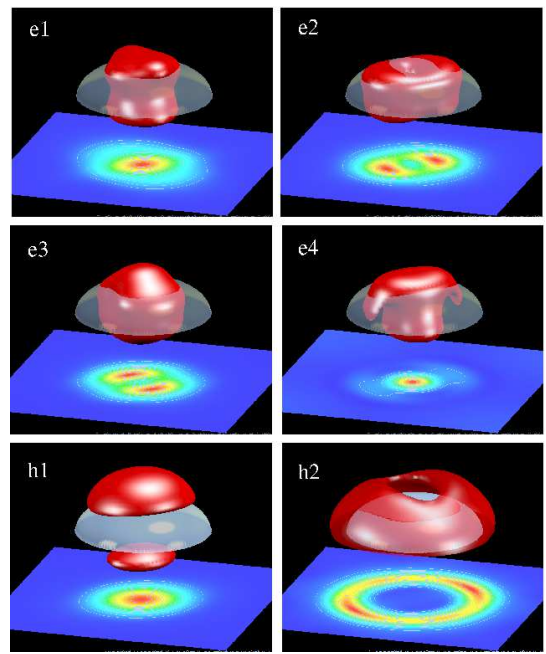}
\caption{COLOR: Wavefunction of the InAs-confined electron states (e1-e4)
and the first two hole states (h1-h2). The transparent lenses indicate
the positions of InAs dots. The isosurface enclose 50 \% of the state
density except for e4 which is only weakly confined. For e4, the isosurface
enclose only about 10 \% of the state density. The contour plot are slices of
the density taken from choosen planes.
  }
\label{fig:states}
\end{figure}

{\it Single-particle states}: 
Diagonalization of the atomistic single-particle pseudopotential Hamiltonian
gives the confined single-particle electron and hole energy levels
shown in Fig. \ref{fig:energy}.
The corresponding wavefunctions for D2 are 
shown in Fig.~\ref{fig:states}.
We see that the four InAs-confined electron states e1, e2, e3, e4 
move {\it down} in 
energy as the dot size increases; whereas the interfacially confined 
(see. Fig.~\ref{fig:states})
hole states h1 and h2 move {\it up} in energy as the size increases. 
For the smallest dot D1, there is a finite, 
single-particle gap of 96 meV between the first confined 
electron and hole states. 
For dot D2, the gap reduces to $\sim$6 meV. 
For the large dot D3, the lowest InAs-confined electron level lies 
{\it below} 
the VBM of the InSb barrier and hybridizes with it.
The single-particle gap is thus negative,
signaling a charge transfer from InSb to InAs.

% Electron states

\begin{figure*}
\includegraphics[width=6.0in,angle=0]{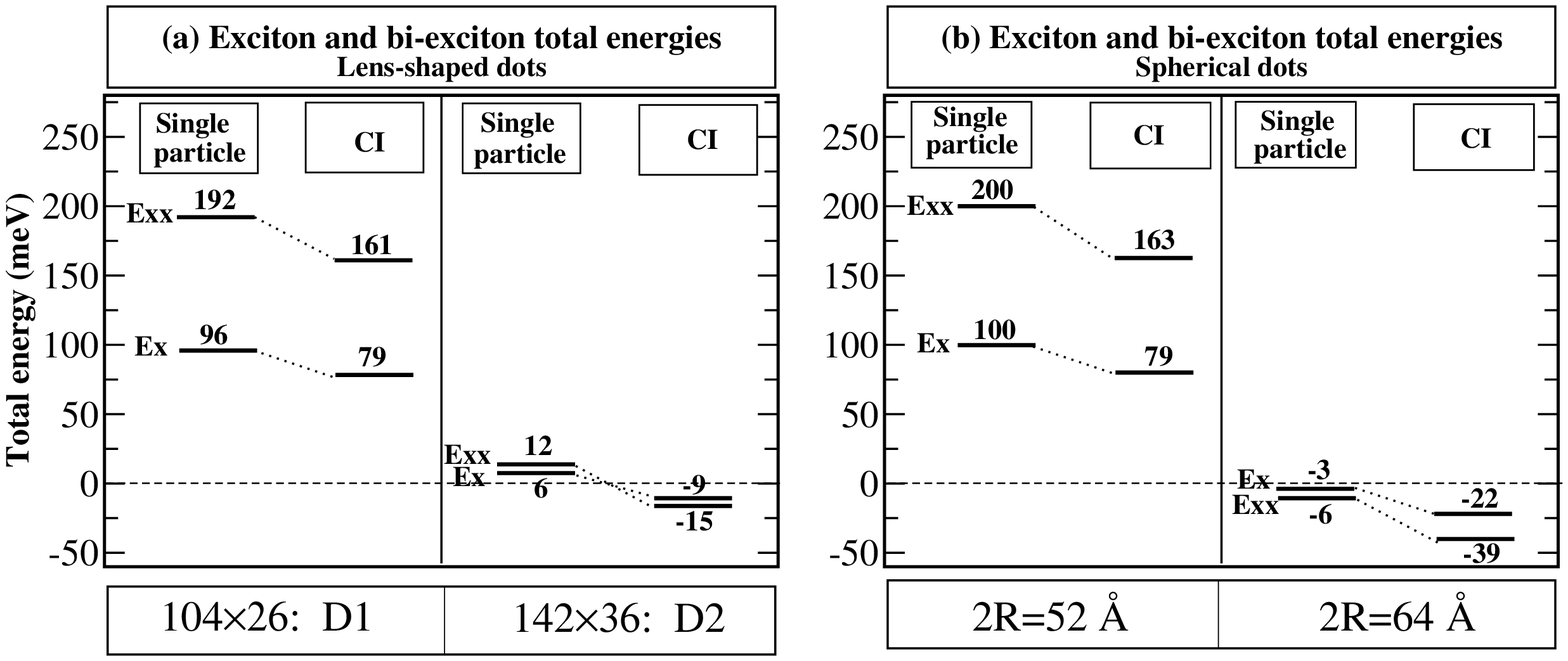}
\caption{ Exciton and Bi-exciton energies of (a) lens-shaped,
and (b) spherical, InAs dots embedded in InSb matrix.}
\label{fig:exciton}
\end{figure*}

{\it Many-body states}:
The energies were carefully converged by increasing the basis
in the many-body expansion. 
For an uncorrelated electron-hole pair, the 
energy is $E_g$, whereas correlation could reduce the energy
of the monoexciton $E_X$ by $\Delta_X \equiv E_g-E_X$. 
Bi-excitons $E_{XX}$ could be
bound with respect to two excitons by $\Delta_{XX} \equiv 2E_X-E_{XX}$. 
The CI energies are shown in Fig.~\ref{fig:exciton}.
The results show that for dot D1,
$E_X$=79 meV  and $E_{XX}$=161 meV both positive i.e, 
correlation energies do not
overcome the single-particle gap. 
However, for dot D2 we find that correlation reduces the energy of the single
electron-hole pair from 6 meV to $E_X=-9$ meV, whereas for 
two electron-hole pairs, 
the reduction is from 12 meV to $E_{XX}=-15$ meV. Thus, both mono- and
bi-excitons are now stable, the bi-exciton being the ground state, lower in
energy than the ``no exciton'' case of a fully occupied valence bands 
and fully empty conduction bands state.   
To see whether the shape of the dot affects these conclusions, 
we performed similar calculations for spherical InAs/InSb
QDs. The critical diameter for the metal-nonmetal transition is now
around 2$R$=64 \AA. The calculated exciton and bi-exciton energies 
are shown in Fig.~\ref{fig:exciton}(b) for two diameters (52 \AA\, and 64 \AA). 
For the 2$R$=64 \AA\, dot, we find an exciton energy of $E_{\rm X}=-22$ meV, and
a bi-exciton energy of $E_{\rm XX}=-39$ meV.
Again the bi-exciton is the
ground state.

{\it Properties of excitonic ground state}:
For lens-shaped dots, the monoexciton binding energy $\Delta_X$ are 17 
and 15 meV for dot D1 and D2 respectively, 
whereas the negative  bi-exciton binding energy of $\Delta_{XX}= -3$ meV for both D1 
and D2, indicating an undbond bi-exciton. 
For the spherical dot with  2$R$=64 \AA\, ,
 $\Delta_X= 19$ meV and $\Delta_{XX}=-5$ meV. The bi-exciton is also unbound. 
The bi-excitonic ground state is
similar to a helium atom in the sense that there are 
two positive and two negative Coulomb bound charges.
These states can be identified as
``excitonic molecule'' states~\cite{littlewood02}.
The calculated permanent electric 
dipole moment 
${\bf p} =\int {\bf {x}} \rho({\bf x}) d^3 x \, $ 
for the exciton and bi-exciton of the lens-shaped dot D2
are 
aligned along (001) direction (growth direction), and amount to
$p_z$=3.58 e\AA\, for the exciton and $p_z$=7.16 e\AA\, for
the bi-exciton. These dipoles could be 
measured experimentally and since the exciton binding energy is so 
high ($\sim$ 15 meV), it can be studied at rather high temperatures.  

Given these excitonic dipoles, 
it is interesting to study the collective behavior of excitons in
the QD arrays due to dipolar interactions.
Recent developments in epitaxial growth made it now possible to 
grow arrays of QDs~\cite{xie95}.
At a  {\it low dot density}, 
where the QDs are far apart, the excitons 
behave like in the single dot  
while couple only weakly via dipole-dipole interactions.
For simplicity, we ignore the fine structure~\cite{bester03}
of the excitons
and assume there can be only one exciton per dot at most
(the results can be easily generalized to the case
allowing two excitons per dot).
The excitonic phase transition can be described 
using an Ising-like Hamiltonian, 
\begin{equation}
H=E_X(\eta) \sum_i n_i + {1\over2}\sum_{i,j} J_{ij}(\eta) n_i\, n_j \, ,
\nonumber
\label{eq:ising}
\end{equation}
where, $n_i$ is the number of excitons on dot $i$, and its
value can be either 0 or 1. 
$E_X(\eta)$ is the formation energy of an 
exciton on a single dot, 
which is a function of $\eta$ being 
dot size, density etc.
The dipole-dipole interaction $J_{ij}$ between the 
exciton on site $i$ and site $j$
is a function of both dot density and array
structures.
We consider here the case where the average interaction 
$\left<J\right>=1/2N\sum_{i,j} J_{ij}(\eta)<0$ (ferromagnetic-like),
where $N$ is the total number of QDs in the array. 
At T=0 K, when $E(\eta)+\left<J\right><$0, all QDs are occupied by excitons,
thus the ground state is an
{\it excitonic state}, accompanied by a
macroscopic dipole moment. When $E(\eta)+\left<J\right> >$0, 
no exciton will form in the QDs lattice,
resulting in the {\it normal} insulting state.
Particular interesting is the quantum critical point, where
$E(\eta)+\left<J\right>$=0. Here the fully occupied excitonic state and
the fully empty state have the same energy, and the state could be
either excitonic or normal, which can be switched by applying an 
electric field. However, at a critical temperature
T$_c$, there are large fluctuations between these two states 
leading to large fluctuations of dipole moment. 
Furthermore, the size distribution and disorder
of the QDs lattice may lead to more complicated 
``Bose glass'' phase~\cite{hertz79,fisher89}.

At {\it medium and high} densities, where the QDs have stronger coupling, 
there could be numerous phases as a function of 
carrier density and mass ratios,
as proposed in Ref.~\onlinecite{littlewood02}.
In these regions, the picture of (bi-)exciton 
on a single QD is not valid anymore.
However, the electron-hole density and mass ratio can be tuned via
the parameters of dots arrays, such as the size and confining potential of
a single dot, the density of dots, and even the structure of the dots
arrays, thus make them very promising for studying the electron-hole phase
diagram.

To conclude, 
our pseudopotential many-body calculations predict
a valence-to-conduction single-particle 
energy level cross-over and a spontaneous formation of  
bi-excitonic ground state at the critical size of 
lens-shaped and spherical 
InAs/InSb dot system.
There are two reasons that make this type-III QD system a
very promising
candidate for studying excitonic ground state. 
First, the electron-hole energy gap is
tunable by the QDs size alone, 
and second, the excitons have much larger
binding energies than in the bulk, or QW due to the 0D confinement
for both electrons and holes.
Experimental studies on such a proposed excitonic ground-state 
are called for.

We thank R. Magri for providing us with the InAs and InSb pseudopotentials.
This work was supported by US DOE-SC-BES-DMS, 
grant no. DEAC36-98-GO10337.


\begin{thebibliography}{26}
\expandafter\ifx\csname natexlab\endcsname\relax\def\natexlab#1{#1}\fi
\expandafter\ifx\csname bibnamefont\endcsname\relax
  \def\bibnamefont#1{#1}\fi
\expandafter\ifx\csname bibfnamefont\endcsname\relax
  \def\bibfnamefont#1{#1}\fi
\expandafter\ifx\csname citenamefont\endcsname\relax
  \def\citenamefont#1{#1}\fi
\expandafter\ifx\csname url\endcsname\relax
  \def\url#1{\texttt{#1}}\fi
\expandafter\ifx\csname urlprefix\endcsname\relax\def\urlprefix{URL }\fi
\providecommand{\bibinfo}[2]{#2}
\providecommand{\eprint}[2][]{\url{#2}}

\bibitem[{\citenamefont{Mott}(1961)}]{mott61}
\bibinfo{author}{\bibfnamefont{N.~F.} \bibnamefont{Mott}},
  \bibinfo{journal}{Phil. \ Mag.} \textbf{\bibinfo{volume}{6}},
  \bibinfo{pages}{287} (\bibinfo{year}{1961}).

\bibitem[{kel()}]{keldysh}
\bibinfo{note}{L. V. Keldysh and Y. V. Kopaev, Fiz. Tverd. Tela {\bf 6}, 2791
  (1964) [ Sov. Solid State {\bf 6}, 2219 (1965)]; L. V. Keldysh and A. N.
  Kozlov, Zh. Eksp. Teor. Fiz, {\bf 54}, 978 (1968) [ Sov. Phys. JETP {\bf 27},
  521 (1968)].}

\bibitem[{\citenamefont{Halperin and Rice}(1968)}]{halperin68}
\bibinfo{author}{\bibfnamefont{B.~I.} \bibnamefont{Halperin}} \bibnamefont{and}
  \bibinfo{author}{\bibfnamefont{T.~M.} \bibnamefont{Rice}},
  \bibinfo{journal}{Rev. \ Mod. \ Phys.} \textbf{\bibinfo{volume}{40}},
  \bibinfo{pages}{755} (\bibinfo{year}{1968}).

\bibitem[{\citenamefont{Littlewood et~al.}(2002)\citenamefont{Littlewood,
  Brown, Eastham, and Szymanska}}]{littlewood02}
\bibinfo{author}{\bibfnamefont{P.~B.} \bibnamefont{Littlewood}},
  \bibinfo{author}{\bibfnamefont{G.~J.} \bibnamefont{Brown}},
  \bibinfo{author}{\bibfnamefont{P.~R.} \bibnamefont{Eastham}},
  \bibnamefont{and} \bibinfo{author}{\bibfnamefont{M.~H.}
  \bibnamefont{Szymanska}}, \bibinfo{journal}{Phys. \ Stat. \ Sol. \ (b)}
  \textbf{\bibinfo{volume}{234}}, \bibinfo{pages}{36} (\bibinfo{year}{2002}).

\bibitem[{\citenamefont{Griffin et~al.}(1995)\citenamefont{Griffin, Snoke, and
  Stringari}}]{bec_book}
\bibinfo{editor}{\bibfnamefont{A.}~\bibnamefont{Griffin}},
  \bibinfo{editor}{\bibfnamefont{D.~W.} \bibnamefont{Snoke}}, \bibnamefont{and}
  \bibinfo{editor}{\bibfnamefont{S.}~\bibnamefont{Stringari}}, eds.,
  \emph{\bibinfo{title}{Bose-Einstein Condensation}}
  (\bibinfo{publisher}{Cambridge university press}, \bibinfo{year}{1995}).

\bibitem[{\citenamefont{Butov et~al.}(2002)\citenamefont{Butov, Lai, Lvanov,
  Gossard, and Chemla}}]{butov02}
\bibinfo{author}{\bibfnamefont{L.~V.} \bibnamefont{Butov}},
  \bibinfo{author}{\bibfnamefont{C.~W.} \bibnamefont{Lai}},
  \bibinfo{author}{\bibfnamefont{A.~L.} \bibnamefont{Lvanov}},
  \bibinfo{author}{\bibfnamefont{A.~C.} \bibnamefont{Gossard}},
  \bibnamefont{and} \bibinfo{author}{\bibnamefont{Chemla}},
  \bibinfo{journal}{Nature} \textbf{\bibinfo{volume}{417}}, \bibinfo{pages}{47}
  (\bibinfo{year}{2002}).

\bibitem[{\citenamefont{Bucher et~al.}(1991)\citenamefont{Bucher, Steiner, and
  Wachter}}]{bucher91}
\bibinfo{author}{\bibfnamefont{B.}~\bibnamefont{Bucher}},
  \bibinfo{author}{\bibfnamefont{P.}~\bibnamefont{Steiner}}, \bibnamefont{and}
  \bibinfo{author}{\bibfnamefont{P.}~\bibnamefont{Wachter}},
  \bibinfo{journal}{Phys. Rev. Lett.} \textbf{\bibinfo{volume}{67}},
  \bibinfo{pages}{2717} (\bibinfo{year}{1991}).

\bibitem[{\citenamefont{Datta et~al.}(1985)\citenamefont{Datta, Melloch, and
  Gunshor}}]{datta85}
\bibinfo{author}{\bibfnamefont{S.}~\bibnamefont{Datta}},
  \bibinfo{author}{\bibfnamefont{M.~R.} \bibnamefont{Melloch}},
  \bibnamefont{and} \bibinfo{author}{\bibfnamefont{R.~L.}
  \bibnamefont{Gunshor}}, \bibinfo{journal}{Phys. Rev. B}
  \textbf{\bibinfo{volume}{32}}, \bibinfo{pages}{2607} (\bibinfo{year}{1985}).

\bibitem[{\citenamefont{Zhu et~al.}(1995)\citenamefont{Zhu, Littlewood,
  Hybertsen, and Rice}}]{zhu95}
\bibinfo{author}{\bibfnamefont{X.}~\bibnamefont{Zhu}},
  \bibinfo{author}{\bibfnamefont{P.~B.} \bibnamefont{Littlewood}},
  \bibinfo{author}{\bibfnamefont{M.~S.} \bibnamefont{Hybertsen}},
  \bibnamefont{and} \bibinfo{author}{\bibfnamefont{T.~M.} \bibnamefont{Rice}},
  \bibinfo{journal}{Phys. Rev. Lett.} \textbf{\bibinfo{volume}{74}},
  \bibinfo{pages}{1633} (\bibinfo{year}{1995}).

\bibitem[{\citenamefont{Magri et~al.}(2000)\citenamefont{Magri, Wang, Zunger,
  Vurgaftman, and Meyer}}]{magri00}
\bibinfo{author}{\bibfnamefont{R.}~\bibnamefont{Magri}},
  \bibinfo{author}{\bibfnamefont{L.~W.} \bibnamefont{Wang}},
  \bibinfo{author}{\bibfnamefont{A.}~\bibnamefont{Zunger}},
  \bibinfo{author}{\bibfnamefont{I.}~\bibnamefont{Vurgaftman}},
  \bibnamefont{and} \bibinfo{author}{\bibfnamefont{J.~R.} \bibnamefont{Meyer}},
  \bibinfo{journal}{Phys. Rev. B} \textbf{\bibinfo{volume}{61}},
  \bibinfo{pages}{10235} (\bibinfo{year}{2000}).

\bibitem[{\citenamefont{Cheng et~al.}(1995)\citenamefont{Cheng, Kono, McCombe,
  Lo, Mitchel, and Stutz}}]{cheng95}
\bibinfo{author}{\bibfnamefont{J.~P.} \bibnamefont{Cheng}},
  \bibinfo{author}{\bibfnamefont{J.}~\bibnamefont{Kono}},
  \bibinfo{author}{\bibfnamefont{B.~D.} \bibnamefont{McCombe}},
  \bibinfo{author}{\bibfnamefont{I.}~\bibnamefont{Lo}},
  \bibinfo{author}{\bibfnamefont{W.~C.} \bibnamefont{Mitchel}},
  \bibnamefont{and} \bibinfo{author}{\bibfnamefont{C.~E.} \bibnamefont{Stutz}},
  \bibinfo{journal}{Phys. Rev. Lett.} \textbf{\bibinfo{volume}{74}},
  \bibinfo{pages}{450} (\bibinfo{year}{1995}).

\bibitem[{\citenamefont{Bimberg et~al.}(1999)\citenamefont{Bimberg, Grundmann,
  and Ledentsov}}]{bimberg_book}
\bibinfo{author}{\bibfnamefont{D.}~\bibnamefont{Bimberg}},
  \bibinfo{author}{\bibfnamefont{M.}~\bibnamefont{Grundmann}},
  \bibnamefont{and} \bibinfo{author}{\bibfnamefont{N.~N.}
  \bibnamefont{Ledentsov}}, \emph{\bibinfo{title}{Quantum Dot
  Heterostructures}} (\bibinfo{publisher}{John Wiley \& Sons},
  \bibinfo{year}{1999}).

\bibitem[{\citenamefont{Magri and Zunger}(2002)}]{magri02}
\bibinfo{author}{\bibfnamefont{R.}~\bibnamefont{Magri}} \bibnamefont{and}
  \bibinfo{author}{\bibfnamefont{A.}~\bibnamefont{Zunger}},
  \bibinfo{journal}{Phys. Rev. B} \textbf{\bibinfo{volume}{65}},
  \bibinfo{pages}{165302} (\bibinfo{year}{2002}).

\bibitem[{\citenamefont{Zunger}(2001)}]{zunger01}
\bibinfo{author}{\bibfnamefont{A.}~\bibnamefont{Zunger}},
  \bibinfo{journal}{phys stat. sol (b)} \textbf{\bibinfo{volume}{224}},
  \bibinfo{pages}{727} (\bibinfo{year}{2001}).

\bibitem[{\citenamefont{Madelung}(1987)}]{lb22a}
\bibinfo{editor}{\bibfnamefont{O.}~\bibnamefont{Madelung}}, ed.,
  \emph{\bibinfo{title}{Landolt-B\"{o}rnstein}}, vol. \bibinfo{volume}{22a}
  (\bibinfo{publisher}{Springer-Verlag Berlin Heidelberg},
  \bibinfo{year}{1987}).

\bibitem[{\citenamefont{Pinczolits et~al.}(1998)\citenamefont{Pinczolits,
  Springholz, and Bauer}}]{pinczolits98}
\bibinfo{author}{\bibfnamefont{M.}~\bibnamefont{Pinczolits}},
  \bibinfo{author}{\bibfnamefont{G.}~\bibnamefont{Springholz}},
  \bibnamefont{and} \bibinfo{author}{\bibfnamefont{G.}~\bibnamefont{Bauer}},
  \bibinfo{journal}{Appl.\ Phys.\ Lett.} \textbf{\bibinfo{volume}{73}},
  \bibinfo{pages}{250} (\bibinfo{year}{1998}).

\bibitem[{\citenamefont{Keating}(1966)}]{keating66}
\bibinfo{author}{\bibfnamefont{P.~N.} \bibnamefont{Keating}},
  \bibinfo{journal}{Phys. Rev.} \textbf{\bibinfo{volume}{145}},
  \bibinfo{pages}{637} (\bibinfo{year}{1966}).

\bibitem[{\citenamefont{Pikus and Bir}(1961)}]{pikus61}
\bibinfo{author}{\bibfnamefont{G.~E.} \bibnamefont{Pikus}} \bibnamefont{and}
  \bibinfo{author}{\bibfnamefont{G.~L.} \bibnamefont{Bir}},
  \bibinfo{journal}{Phys. \ Rev. \ Lett.} \textbf{\bibinfo{volume}{6}},
  \bibinfo{pages}{103} (\bibinfo{year}{1961}).

\bibitem[{\citenamefont{Wang and Zunger}(1999)}]{wang99b}
\bibinfo{author}{\bibfnamefont{L.-W.} \bibnamefont{Wang}} \bibnamefont{and}
  \bibinfo{author}{\bibfnamefont{A.}~\bibnamefont{Zunger}},
  \bibinfo{journal}{Phys.\ Rev.\ B} \textbf{\bibinfo{volume}{59}},
  \bibinfo{pages}{15806} (\bibinfo{year}{1999}).

\bibitem[{\citenamefont{Bester et~al.}(2003)\citenamefont{Bester, Nair, and
  Zunger}}]{bester03}
\bibinfo{author}{\bibfnamefont{G.}~\bibnamefont{Bester}},
  \bibinfo{author}{\bibfnamefont{S.~V.} \bibnamefont{Nair}}, \bibnamefont{and}
  \bibinfo{author}{\bibfnamefont{A.}~\bibnamefont{Zunger}},
  \bibinfo{journal}{Phys. Rev. B} \textbf{\bibinfo{volume}{67}},
  \bibinfo{pages}{161306} (\bibinfo{year}{2003}).

\bibitem[{ins()}]{insb_pp_note}
\bibinfo{note}{The InSb pseudopotentials are revised to reproduce correct
  band-offset of InSb VBM relative to GaSb, changing it from 40 meV below the
  GaSb VBM to 34 meV above one.}

\bibitem[{\citenamefont{Franceschetti et~al.}(1999)\citenamefont{Franceschetti,
  Fu, Wang, and Zunger}}]{franceschetti99}
\bibinfo{author}{\bibfnamefont{A.}~\bibnamefont{Franceschetti}},
  \bibinfo{author}{\bibfnamefont{H.}~\bibnamefont{Fu}},
  \bibinfo{author}{\bibfnamefont{L.-W.} \bibnamefont{Wang}}, \bibnamefont{and}
  \bibinfo{author}{\bibfnamefont{A.}~\bibnamefont{Zunger}},
  \bibinfo{journal}{Phys.\ Rev.\ B} \textbf{\bibinfo{volume}{60}},
  \bibinfo{pages}{1819} (\bibinfo{year}{1999}).

\bibitem[{\citenamefont{Onida et~al.}(2002)\citenamefont{Onida, Reining, and
  Rubio}}]{onida02}
\bibinfo{author}{\bibfnamefont{G.}~\bibnamefont{Onida}},
  \bibinfo{author}{\bibfnamefont{L.}~\bibnamefont{Reining}}, \bibnamefont{and}
  \bibinfo{author}{\bibfnamefont{A.}~\bibnamefont{Rubio}},
  \bibinfo{journal}{Rev. Mod. Phys.} \textbf{\bibinfo{volume}{74}},
  \bibinfo{pages}{601} (\bibinfo{year}{2002}).

\bibitem[{\citenamefont{Xie et~al.}(1995)\citenamefont{Xie, Madhukar, Chen, and
  Kobayashi}}]{xie95}
\bibinfo{author}{\bibfnamefont{Q.}~\bibnamefont{Xie}},
  \bibinfo{author}{\bibfnamefont{A.}~\bibnamefont{Madhukar}},
  \bibinfo{author}{\bibfnamefont{P.}~\bibnamefont{Chen}}, \bibnamefont{and}
  \bibinfo{author}{\bibfnamefont{N.}~\bibnamefont{Kobayashi}},
  \bibinfo{journal}{Phys. Rev. Lett.} \textbf{\bibinfo{volume}{75}},
  \bibinfo{pages}{2542} (\bibinfo{year}{1995}).

\bibitem[{\citenamefont{Hertz et~al.}(1979)\citenamefont{Hertz, Fleishman, and
  Anderson}}]{hertz79}
\bibinfo{author}{\bibfnamefont{J.~A.} \bibnamefont{Hertz}},
  \bibinfo{author}{\bibfnamefont{L.}~\bibnamefont{Fleishman}},
  \bibnamefont{and} \bibinfo{author}{\bibfnamefont{P.~W.}
  \bibnamefont{Anderson}}, \bibinfo{journal}{Phys. Rev. Lett.}
  \textbf{\bibinfo{volume}{43}}, \bibinfo{pages}{942} (\bibinfo{year}{1979}).

\bibitem[{\citenamefont{Fisher et~al.}(1989)\citenamefont{Fisher, Weichman,
  Grinstein, and Fisher}}]{fisher89}
\bibinfo{author}{\bibfnamefont{M.~P.} \bibnamefont{Fisher}},
  \bibinfo{author}{\bibfnamefont{P.~B.} \bibnamefont{Weichman}},
  \bibinfo{author}{\bibfnamefont{G.}~\bibnamefont{Grinstein}},
  \bibnamefont{and} \bibinfo{author}{\bibfnamefont{D.~S.}
  \bibnamefont{Fisher}}, \bibinfo{journal}{Phys. Rev. B}
  \textbf{\bibinfo{volume}{40}}, \bibinfo{pages}{546} (\bibinfo{year}{1989}).

\end{thebibliography}
\end{document}